\documentclass[letterpaper,twocolumn,10pt]{article}
\usepackage{usenix}

\usepackage{tikz}
\usepackage{amsmath}
\usepackage[compact]{titlesec}
\usepackage{graphicx}
\usepackage{textcomp}
\usepackage{xspace}
\usepackage{hyperref}
\usepackage{xurl}
\usepackage{subcaption}
\usepackage{algorithm, algpseudocode}
\usepackage[normalem]{ulem}

\newcommand{\ray}[1]{#1}
\newcommand{\new}[1]{#1}

\newtheorem{definition}{Definition}

\newcommand{\sys}{\textsc{Chiron}\xspace}

\begin{document}
%-------------------------------------------------------------------------------

%don't want date printed
\date{}

\title{\Large \bf \sys: Accelerating Node Synchronization without Security Trade-offs in Distributed Ledgers}

% \sys: Supercharging Parallel Smart Contract Execution

\author{
{\rm Ray Neiheiser}\\
ISTA
\and
{\rm Giannis Alexopoulos}\\
TU Wien
\and
{\rm Arman Babaei}\\
ISTA
\and
{\rm Marios Kogias}\\
Imperial College London
\and
{\rm Eleftherios Kokoris Kogias}\\
Mystenlabs \& ISTA
}

\maketitle

%-------------------------------------------------------------------------------
\begin{abstract}
%-------------------------------------------------------------------------------

Parallel smart contract execution engines are a promising solution for scaling blockchain throughput. However, the actual speed-up is constrained by the characteristics of the underlying workload.
Furthermore, rising hardware demands for validators and full nodes introduce practical barriers to participation, thereby increasing the risk of centralization. This is especially staggering, given that already more than 25\% of Ethereum nodes are unable to keep up at much lower hardware requirements. As a result, struggling full nodes and validators often catch up by synchronizing from signed checkpoints, weakening the underlying trust assumptions.

In response to these challenges, this paper introduces \sys, a system designed to extract execution hints to accelerate struggling full nodes and validators. Notably, \sys achieves this without weakening the trust assumptions and without introducing overhead on the critical path of consensus. Our evaluation results demonstrate 
that stragglers can execute blocks over 300\% faster, effectively addressing the gap between theoretical research and practical deployment. We evaluate this with the help of realistic blockchain benchmarks derived from a comprehensive analysis of real-world workloads, constituting an independent contribution.

%The introduction of modularity in blockchain architectures has led to substantial performance improvements. Separating the consensus and execution layers enables concurrent block dissemination, moving the performance bottleneck to transaction processing.

%As a result, approaches emerged to take advantage of multi-core architectures and scale execution through parallel transaction processing. However, this also significantly increases the resource requirements for validators and full nodes, risking centralization.
%However, existing solutions to allow weaker nodes and full nodes to catch up to the system relax the security guarantees compared to the initial Bitcoin vision.

%In this work, we present \sys where hints are extracted from the execution results that can then be used to speed up straggling validators or full nodes without having to relax the security model. Our experiments show, that depending on the workload, this can speed up parallel execution by up to 30\%.
\end{abstract}

\section{Introduction}
\label{sec:introduction}

Recent research efforts paved the way for visa-level throughput in Byzantine Fault-Tolerant Consensus~\cite{narwahl,dumbo,bullshark,mysticeti,kauri}, shifting the performance of smart contract execution into focus. This is particularly relevant as many blockchains, such as Ethereum~\cite{ethereum}, still execute transactions sequentially, not taking advantage of multi-core architectures.

Recognizing this challenge, research on parallel execution engines has gained popularity~\cite{blockstm,polygonupdate,10806617}, enabling transactions to be executed in parallel rather than sequentially. Furthermore, in practice (e.g., Solana, Sui, Aptos~\cite{solana,sui,aptos}), parallel execution is often decoupled from consensus, improving throughput~\cite{narwahl} but yielding a \emph{dirty ledger}~\cite{dirtyledger} in which some included transactions are later discarded.
% \ga{I would replace the following sentence with something along the lines of: This modular system design allows one to focus on each layer independently ...}.
% Although this results in a dirty ledger~\cite{dirtyledger} where transactions, even though included in a block, could still be aborted during execution, this approach has been shown to significantly improve throughput~\cite{narwahl}.

Parallel execution engines can be broadly divided into two categories: \textit{Optimistic} and \textit{Guided} execution engines~\cite{blockstm,solana,ethspeculative,ethforward}. Guided execution, as used in Solana~\cite{solana} and Sui~\cite{sui}, relies on an exhaustive set of resource addresses, often referred to as hints, that clients have to send alongside the transaction. When scheduling the transactions for execution they account for potential read-write conflicts, guaranteeing at the application level that no conflicts may arise. If a transaction fails to declare all its dependencies, the execution engine detects the out-of-bounds access and aborts the transaction. In the database context, this is comparable to pessimistic approaches, where locking is used to prevent conflicts.

Meanwhile, optimistic execution engines, as in Aptos~\cite{aptos,blockstm}, optimistically execute transactions in parallel, detect conflicts as they arise, and re-execute transactions when necessary. While, in the worst case, optimistic execution engines may have a high re-execution overhead, they simplify application development, as applications do not have to provide execution hints, allowing an easier integration into existing blockchains where the concept of execution hints does not exist~\cite{polygonupdate}.

However, the speed-up achievable through optimistic execution in real-world deployments remains unclear. For instance, the Polygon team's effort to implement Block-STM resulted in only a small improvement over sequential execution~\cite{polygonupdate}.

Inspired by this, as a first contribution, we identify and close a clear gap between theoretical research and practical deployments; the lack of realistic blockchain workloads that capture the type and frequency of data dependencies and contention. 
In our analysis, we identify high levels of contention that significantly affect the effectiveness of the execution engine.

Beyond their unclear practical impact, blockchains that support parallel transaction execution have exceedingly high hardware requirements that might be disproportionate to the practical speed-up they deliver. This raises centralization concerns as participation becomes increasingly costly, and an increasing fraction of nodes \emph{cannot keep up} with verification. This is particularly staggering given that even in blockchains with significantly lower hardware requirements and throughput, such as Ethereum, more than a quarter of all nodes are unable to keep up as of the time of writing~\cite{nodewatch}.

When nodes lag behind, the consequences are immediate: full nodes respond to clients only after significant delays, and validators may miss their proposal slots or propose a significant number of stale transactions, reducing the goodput of the system. Furthermore, in the presence of faulty nodes, the system might stall until sufficient correct validators catch up to the head of the chain.

In practice, lagging full nodes and validators are under pressure to catch up not by reauditing and reexecuting the full history, but by skipping ahead via signed checkpoints. While this accelerates synchronization, only a smaller subset of active nodes actually verifies the validity of the blocks in the chain. This behavior is compatible with standard BFT assumptions and, by itself, does not threaten safety as long as the fraction of faulty nodes remains below the tolerated threshold.

However, recent work~\cite{recoveryprotocols,roughgarden} has relaxed this assumption with the help of recovery models, where systems can recover from periods of Byzantine majority while maintaining economic safety, i.e., client losses can be compensated from slashed stake.
To ensure economic safety in these settings, clients' potential losses must be bounded. Consequently, several works assume that all validators and full nodes always verify block validity to limit the damage of potential attacks under an adversarial majority~\cite{cobra,stakesure}.
This assumption also aligns with Bitcoin’s original vision for validators and full nodes~\cite{bitcoin,centralization,bitcoinfullnodes}. We refer to this property as unconditional validity.

\ray{Although zero-knowledge proofs could theoretically be used to attest to the correctness of execution outputs to aid recovery, their generation imposes a substantial computational overhead which renders this approach impractical.}

In this paper, we propose \sys, a framework that helps struggling validators and full nodes, hereafter denoted as stragglers, catch up in blockchains that deploy optimistic execution engines, for example, Aptos~\cite{aptos}, Monad~\cite{monad} or Sei~\cite{sei}, without sacrificing unconditional validity. \ray{Notably, \sys achieves this with minimal overhead and without relying on sophisticated cryptographic mechanisms.}
\sys achieves this and bridges the gap between optimistic and guided execution engines by extracting accurate hints from the execution results and providing them to stragglers.
Furthermore, \sys does not rely on the accuracy of hints for safety, does not affect the liveness of the underlying protocol and does not introduce any overhead on the critical path of consensus. This reestablishes the initial security vision of Bitcoin with minimal overhead. Our evaluation of \sys shows that stragglers can, depending on the workload, execute blocks over 300\% faster.

In summary, we provide the following contributions:

\begin{itemize}
    %\marios{We use the conclusion of this analaysis regarding the smart contract dependencies in the design of \sys.}
    %\item We propose \sys, an extension of BlockSTM that uses hints to build an optimal execution schedule but does not rely on it for safety.
    \item We perform an analysis of popular real-world blockchain applications and develop a microbenchmark for parallel smart contract execution engines based on the findings.
    \item We propose \sys, a framework to enable stragglers to catch up through guided execution without relaxing security guarantees.
    \item We evaluate \sys under the proposed microbenchmark and show a speed-up of over 300\% compared to optimistic parallel execution.
    %\item We provide a fresh view and reevaluation of the reliance on certain security thresholds which present a deterioration of the security guarantees compared to the original Bitcoin design.
\end{itemize}

%The paper is organized as follows. First, in Section~\ref{sec:systemmodel} we outline the system model before we introduce \sys in Section~\ref{sec:overview}. Following that, in Section~\ref{sec:workloads}, we discuss the extracted blockchain workloads in detail and explain how we construct a portable benchmark based on the workloads. Following that, in Section~\ref{sec:evaluation} we evaluate \sys under the proposed benchmarks. Finally, in Section~\ref{sec:discussion} we discuss the related and future work before we conclude the paper in Section~\ref{sec:conclusion}.

\section{Blockchain Workloads}
\label{sec:workloads}

Most benchmarks in academia and industry that evaluate parallel smart contract execution engines either generate random, artificial peer-to-peer transfers~\cite{blockstm} with uniform distributions or apply non-blockchain-based workloads~\cite{diablo} such as Uber's, Youtube's, or Twitter's. However, they poorly reflect the workloads encountered by production blockchains, thereby failing to highlight the shortcomings of existing parallel smart contract execution engines.
While some works~\cite{polygonupdate} evaluate the performance of the execution engine by re-executing past transaction history, this approach has two important limitations. 
First, these workloads are confined to their respective ecosystems and cannot be easily ported to other blockchains, making it difficult to compare the performance of competing approaches across different ecosystems.  
Moreover, when replaying transaction histories from ecosystems that do not natively support parallel execution, there may be significantly more conflicts than necessary, as the smart contracts were not designed with concurrency in mind.

In contrast, for our workloads, we extracted the essential points of contention from a range of application settings and constructed simple smart contracts, enabling the usage of our microbenchmark across ecosystem and framework boundaries.

In this section, we analyze the transaction history of popular blockchains and blockchain applications. The purpose of this analysis is twofold:
First, to identify acceleration opportunities for \sys through careful execution scheduling based on hints, compared to existing optimistic parallel execution engines.
Second, to provide a microbenchmark consisting of a set of realistic and versatile blockchain workloads for our evaluation and the wider community.
We aim to design a microbenchmark that accurately reflects real-world workloads, to allow estimating the average throughput that can be sustained in practice and reveal potential bottlenecks.

\paragraph*{\textbf{Overview}}
% \subsubsection*{Overview}

First, we want to identify application settings that are commonly encountered in most blockchain ecosystems.
At the time of writing, CoinMarketCap~\cite{coinmarketcap} lists over 10,000 tokens and NFTs, as well as nearly 500 registered decentralized exchanges. This creates a solid foundation for workloads that most blockchains have in common; Peer-to-Peer Transactions, NFT Minting, and Decentralized Exchange Trading. Additionally, to account for generic smart contract interactions, we also want a Mixed workload that can capture the average accesses and computational complexity that smart contracts and their interactions exhibit.

These workloads cover a wide range of execution characteristics, from heavy contention and complex contract interactions to simple Peer-to-peer transactions, allowing for a more comprehensive evaluation of execution engines and their ability to handle the demands in a real-world blockchain setting.
In all workloads, contention stems from two or more transactions accessing the same resource, where resources could be balances, counters, vectors, or other data structures in smart contracts. We call a resource that appears in a large percentage of transactions a \textit{hot} resource.

To make our microbenchmark easily portable across ecosystem boundaries, we aim to design smart contracts that are as simple as possible to focus on the essential elements of contention.  
As a first step, we analyzed popular example smart contracts, including NFTs (e.g., BAYC), Peer-to-Peer Transfers (e.g., Aptos and Ethereum Transfers), and Decentralized Exchanges (e.g., Uniswap) and identified the following essential sources of contention for the different workloads:  

First, NFT smart contracts maintain a counter that is incremented with each minting transaction. Consequently, transactions attempting to mint the same NFT cannot be executed in parallel, as the index determines the resulting NFT and enforces limits on the number of NFTs that can be minted at a given contract. As this enforces a deterministic order of mint operations, this can not easily be parallelized.

Next, for Peer-to-Peer transactions in the account model or the \textit{Object Model}~\cite{move}, two transactions that interact with the same user balance create a conflict. For example, if user A receives tokens from user B while simultaneously sending tokens to user C, these transactions conflict due to the balance update for user A.  
While two transactions simultaneously depositing tokens into the same balance could, in theory, be partially parallelized using an atomic increment operation, to the best of our knowledge, no blockchain employing the parallel execution model has implemented this optimization for Peer-to-Peer transactions at the time of writing.  

Finally, in the case of decentralized exchanges, Uniswap employs the automated market maker model and uses liquidity pools to optimize efficiency. Conflicts arise when transactions interact with the same liquidity pool. Thus, two transactions attempting to trade the same coin pair will conflict.

\paragraph*{\textbf{Data Extraction}}
% \subsubsection*{Data Extraction}

Next, having identified the essential points of conflict, we require real-world data to inform the design of our microbenchmark.
Therefore, we analyzed blockchain transaction histories of 2022.
For NFT minting and Peer-to-Peer transactions, we analyzed the transaction history of Ethereum, as it hosts some of the most popular NFTs and tokens, and for DEX workloads, we obtained trading data from Uniswap, one of the leading decentralized exchanges.
Finally, for the mixed workload, we chose Solana for several reasons. First, at the time of writing, Solana is the most popular platform that supports parallel execution. Furthermore, Solana transactions include hints about the resources they access, allowing us to analyze the access distribution of resources. In addition to that, we use the gas consumption of the individual transaction as a measurement of their execution complexity.  

As we aim to test the throughput limits of parallel execution engines we want to be able to build large blocks of up to 10000 transactions. However, blocks on Ethereum and Solana only hold, on average, several hundred non-system transactions per block. Furthermore, after filtering by application this reduces the set of transactions significantly.
Therefore, we computed the average access distribution across 1000 blocks for the NFT, P2P, and Mixed workloads. This way we obtain sufficient transactions of each workload type while also taking the time dependency of transactions into account.
For the DEX workload, we conducted a day-by-day assessment and divided it into two categories: The average daily volume and the average volume of the thirty most highly contended days of the year.
The algorithms for constructing the smart contracts can be found in Appendix~\ref{app:alg}.

In the following, we describe the workloads in detail:

\begin{figure*}[t]
\centering
\includegraphics[width=\linewidth]{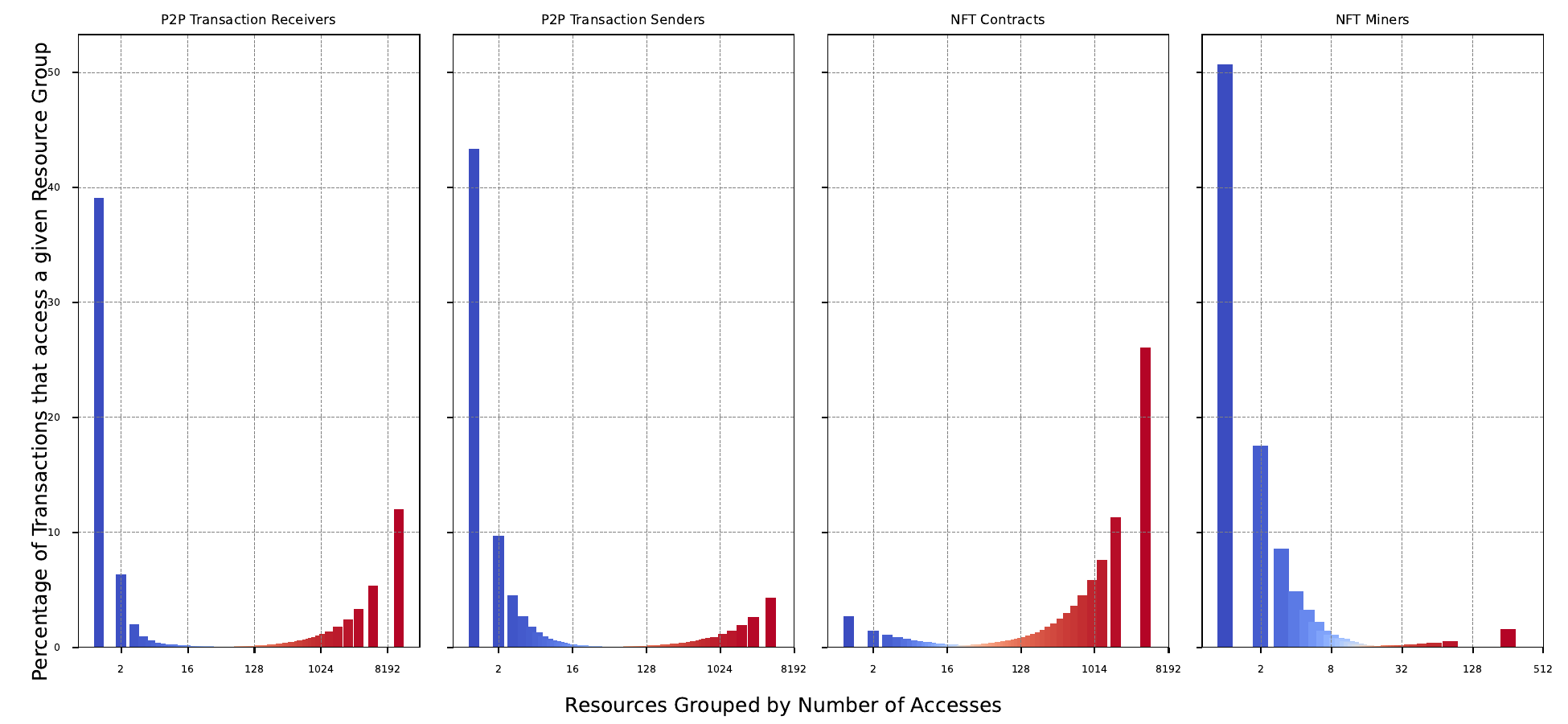} 
\caption{Distribution of P2P and NFT Resources}
\label{fig:ethworkloads2}
\end{figure*}

\begin{figure}[ht]
\centering
\includegraphics[width=\linewidth]{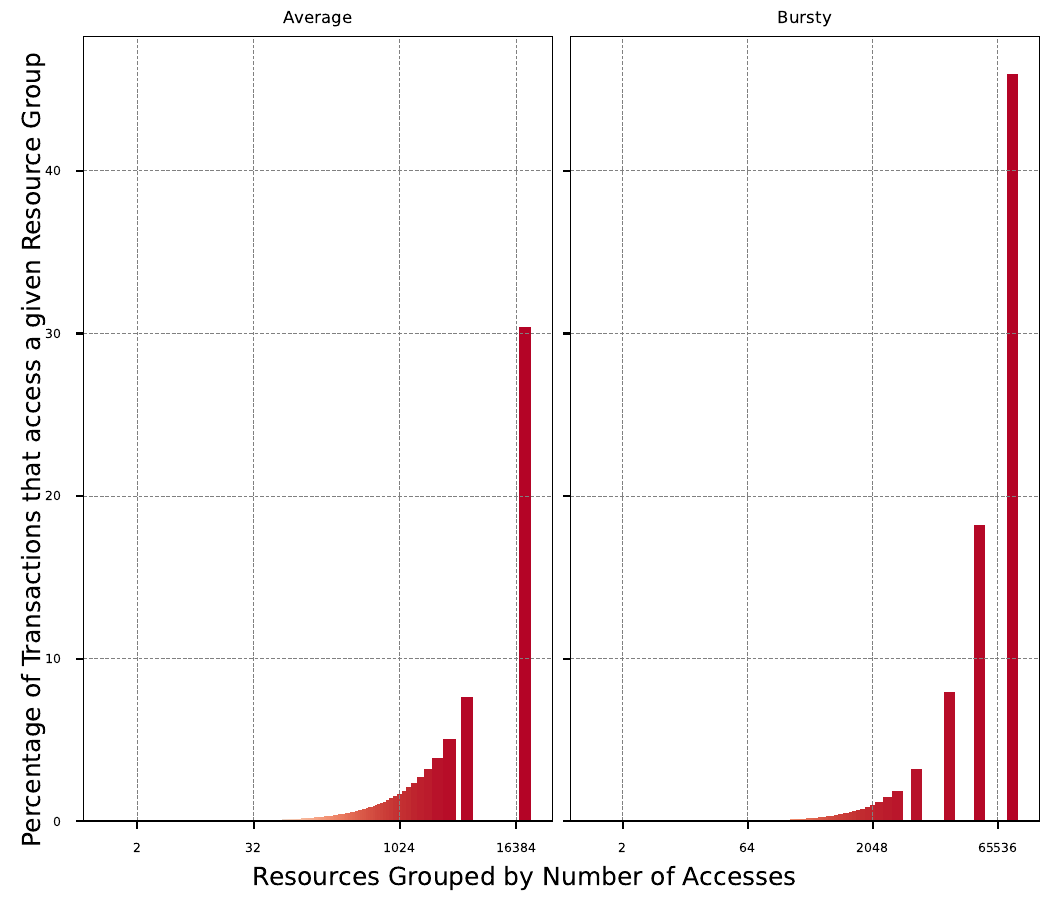}
\caption{Distribution of Resources on Uniswap}
\label{fig:uniswapworkload2}
\end{figure}

\paragraph{Peer-to-Peer Transaction Workload}

First, the \textit{Peer-to-Peer Transaction} workload where, instead of assuming a uniform distribution, we measured the distribution of senders and receivers of payment transactions on Ethereum throughout 2022.
Conflicts in this workload arise when two transactions have overlapping participants (e.g. same receiver).
Figure~\ref{fig:ethworkloads2} depicts the popularity of the hottest resources in this workload. On the x-axis we group resources by their popularity, where on the leftmost side we have the least popular resources that were only accessed once within each measurement period on the right side the most popular resource that was accessed thousands of times.
Meanwhile, the y-axis shows how often each resource is accessed compared to all other resources in percent.
The hottest resource, in this case, the most popular receiver, appears in over 10\% of transactions. Furthermore, the three hottest resources combined appear in around 20\% of all transactions, significantly limiting the potential parallelism.

\paragraph{NFT Workload}

Next, the \textit{NFT Minting} workload is derived from the NFT minting activity on Ethereum in 2022. 
In the case of NFT minting, we expect each transaction that mints the same NFT to conflict due to the NFT index that is incremented with each transaction.
Compared to the P2P workload, as shown in Figure~\ref{fig:ethworkloads2}, the hottest resource already appears in around 25\% of all transactions, while the top three resources appear in almost half of all transactions, showing severely more contention than the P2P workload.

\paragraph{Decentralized Exchange Workload}

In the context of decentralized exchanges, we created two \textit{DEX Workloads} for which we gathered data on the daily distribution of different trading pairs on Uniswap~\cite{uniswap} throughout 2022.
First, an \textit{Average DEX Workload} derived from the annual average. Furthermore, as we observed a large variance in the daily distribution of trading behavior, we created a \textit{Bursty DEX Workload} that we computed based on the average distribution of the thirty most contended days.
In Uniswap DEX smart contracts, transactions trading the same coin-pair touches the same liquidity pool and as such conflict with each other.

As shown in Figure~\ref{fig:uniswapworkload2}, in both cases, the hottest resources surpass the contention of previous workloads, accounting for over 30\% in the Average DEX workload and over 45\% in the Bursty DEX workload. In the context of the three hottest resources, the Average DEX workload shows slightly less contention compared to the NFT workload, while in the Bursty DEX workload, the three hottest resources account for up to 70\% of all transactions.

\paragraph{Mixed Workload}

\begin{figure*}[ht]
\centering
\includegraphics[width=\linewidth]{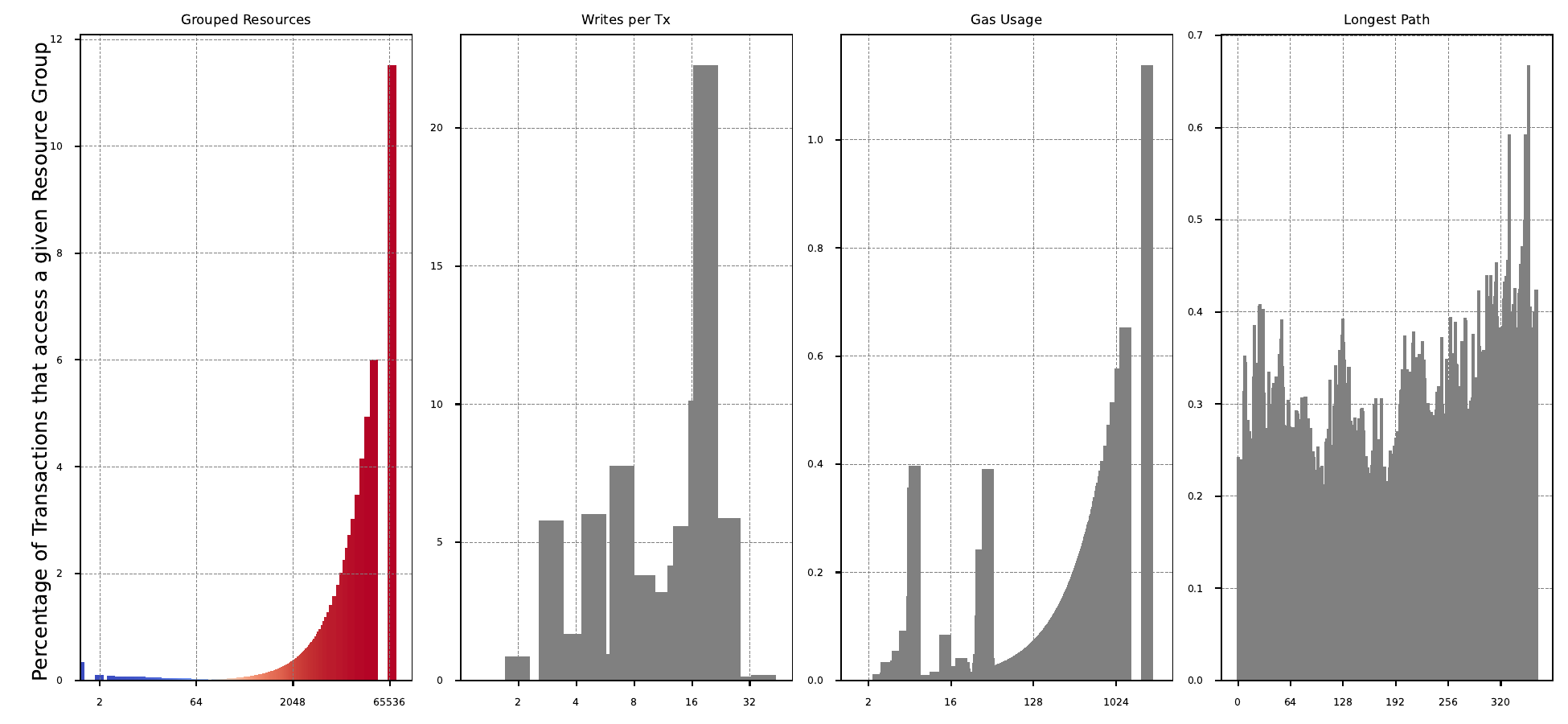}
\caption{Mixed Workload}
\label{fig:mixedworkload}
\end{figure*}

Finally, for the \textit{Mixed Workload}, we extracted the write sets of Solana transactions and their corresponding gas expenditures.
This workload is the most complex among the four, as it involves varying the length of the write-set, the access distribution of resources within the write-set, and the transaction runtime. Due to the large number of blocks produced daily on Solana, we sampled 1000 blocks per day at regular intervals throughout 2022 and discarded the system maintenance transactions. %As Solana requires clients to declare the resources a transaction accesses during execution, we obtained the declared resources from the respective transactions. 
The key results are shown in Figure~\ref{fig:mixedworkload}. 
Analogous to the other workloads, we observe that a small number of resources make up a large percentage of the write accesses, where the three hottest resources are accessed by over 20\% of transactions. Furthermore, we observe that most transactions access several resources and that the execution times are well distributed.
The access distribution of the resources results in sequential paths taking up between 20\% and almost 70\% of all transactions with an average of around 30\%. This is the case, as many transactions access several popular resources.
This also provides insight into the variance of the workload on Solana, where at times, popular applications cause  70\% of transactions in a given interval requiring sequential execution.

This was calculated by comparing the total gas consumption over 1000 blocks with the combined gas cost of the longest path of dependent transactions within the same period.
To make sure that the sequential path in the benchmark approximates what we observed in the data, we adjusted the workload generation code such that the resource distribution, number of writes, and transaction length, on average result in a sequential path of around 30\%.
Therefore, the mixed workload has transactions with varying numbers of writes and varying computational complexity.

\paragraph*{Workload Algorithms}
\label{app:alg}

\begin{algorithm}[ht!]
\scriptsize
\centering
\caption{\textsl{P2P Smart Contract}}
\label{algo:p2ptx}
\begin{algorithmic}[1]
\State{$resourcetable \gets \emptyset$}

\Procedure{accesstwo}{$addr1, addr2$}
\For{$i=1 \to R$}
    \State{$resourcetable[addr1]+=1$}
    \State{$resourcetable[addr2]+=1$}
\EndFor
\EndProcedure
\end{algorithmic}
\end{algorithm}

We created simple smart contracts for our benchmark, that can be ported trivially to any smart contract language. Algorithm~\ref{algo:p2ptx} shows the pseudocode for the P2P smart contract. The smart contract holds a resource table, and each transaction accesses two resources (i.e. sender and receiver balance).
We run the table assignment in a loop of $R$ iterations to simulate a realistic runtime, given that in most blockchain ecosystems there is a comprehensive number of checks (reads and writes) for each P2P transaction. This parameter can be obtained by comparing the execution time of the simplified smart contract with the runtime of a regular transaction in the given ecosystem.

\begin{algorithm}[ht!]
\scriptsize
\centering
\caption{\textsl{Single Resource Smart Contract}}
\label{algo:nftdex}
\begin{algorithmic}[1]
\State{$resourcetable \gets \emptyset$}

\Procedure{accessone}{$addr1$}
\For{$i=1 \to R$}
    \State{$resourcetable[addr1]+=1$}
\EndFor
\EndProcedure
\end{algorithmic}
\end{algorithm}

Next, the pseudocode for the NFT and DEX workloads is displayed in Algorithm~\ref{algo:nftdex}. Each transaction accesses a single resource in the table and increments the value $R$ times, analog to the P2P workload to simulate a realistic runtime complexity.

\begin{algorithm}[ht!]
\scriptsize
\centering
\caption{\textsl{Multi Resource Complex Runtime Smart Contract}}
\label{algo:mixed}
\begin{algorithmic}[1]
\State{$resourcetable \gets \emptyset$}

\Procedure{accessn}{$complexity, setofaddresses$}
\State{$i \gets 0$}
\For{\textbf{all} $addr \in setofaddresses$}
    \State{$i+=1$}
    \State{$resourcetable[addr]+=1$}
\EndFor

\For{$j=i \to complexity$}
    \State{$setofaddresses[j\%|setofaddresses|]+=1$}
\EndFor
\EndProcedure
\end{algorithmic}
\end{algorithm}

Finally, the pseudocode for the mixed workloads smart contract is outlined in Algorithm~\ref{algo:mixed}.
First, in a loop, we access the resource table and increment the value of each resource in the set of addresses at least once.
Next, following the complexity parameter, we iterate an additional $complexity - i$ times and access the addresses in the set of addresses uniformly.

\paragraph*{\textbf{Summary}}

With the help of the access distribution, we sample transactions from the respective workloads to generate the microbenchmarks. For example, if 10\% of transactions in the workload access the same resource and the remaining 90\% access independent resources, the microbenchmark reflects this distribution. Given the 10\% example, the probability of having two transactions access the same resource back-to-back in the block is 1\%. Note that, given our workload analysis, which shows that independent users regularly access the same resource, we find that this correctly reflects the average distribution even from a fine-grained perspective.

For each workload, we constructed probability distributions in the shape of $[1,1,1,1,10,100]$, where each entry corresponds to a specific resource, and the value indicates the weight of that resource being accessed. To generate the workload, resources are selected iteratively based on their weighted probabilities within the distribution\footnote{\url{https://anonymous.4open.science/r/chain-execution-benchmark-1BFB}. }.

In summary, our workload analysis confirms that except for the peer-to-peer workload, all workloads are highly contended, validating our initial claim. Due to this, naively executing smart contracts in parallel will lead to high abort rates. 
This leaves an opening for \sys to leverage hints to speed up stragglers.
In the remainder of this paper, we discuss the design of \sys that allows maximizing parallelism in this context.

\section{System Model}
\label{sec:systemmodel}

Before delving into the design of \sys, we first explore the underlying system model.
% upon which we construct \sys.
We assume the existence of a set of $N$ server processes $p_1, p_2, ..., p_N$ and a set of $I$ client processes $c_1, c_2, ..., c_I$ communicating over a peer-to-peer network where clients and servers are identified with the help of asymmetric key pairs~\cite{bitcoin} and entities prove their identity by signing their respective transactions and messages. Furthermore, clients and servers communicate over perfect point-to-point channels achieved through mechanisms for message retransmissions, ordering, and deduplication. Thus, if a process $p_i$ sends a message $m_{ij}$ to process $p_j$, $p_j$ eventually receives $m_{ij}$. 

To deal with network failures, we assume that the network follows a partial synchrony model based on~\cite{dls}. While, during periods of asynchrony, messages may be delayed for an arbitrary amount of time, we assume the existence of regular periods of stability, after some \textit{Global Stabilization Time} (GST). During these periods, messages passed between two correct processes arrive within a known bound $\delta$.

% Analogous to most real-world deployments of parallel execution engines~\cite{solana,sui,aptos}, execution, and consensus are split into modular layers that run in parallel \ga{"Parallel" execution engines that run in "parallel" - I would slightly rephrase, it might be confusing to the reader. maybe say parallel execution engines are mostly deployed in systems with modular design etc...} such that execution does not happen on the critical path of consensus.

Following the trend in real-world deployments of parallel execution engines~\cite{solana,sui,aptos}, we consider a \emph{modular} blockchain architecture that \emph{decouples} execution from consensus. The two layers progress concurrently, so execution is not on the critical path of consensus. For simplicity\footnote{In the general case, a server may choose to assume only one of the two roles i.e., either validator or executor.}, we assume all server processes \(p_1,\ldots,p_N\) take on two roles: \emph{validators}, which run the consensus protocol and deliver to every server an identical, totally ordered block sequence \(B_1,B_2,\ldots,B_n\) and \emph{executors}, which consume this block stream and deterministically apply transactions to local state, producing state updates. Client processes may either act as executors (\emph{active clients}) or receive state updates through other nodes (\emph{passive clients}).

Because execution is not on the critical path of consensus, client transactions cannot be fully validated prior to inclusion. We therefore model the consensus output as a \emph{dirty ledger}~\cite{dirtyledger,executing-over-dirty}, \(\mathcal{L}_i := [tx_1,\ldots,tx_i]\), in which the blockchain may contain invalid transactions (e.g., insufficient funds or gas). Nonetheless, as long as executors receive the same blocks in the same order (guaranteed by consensus) and execution is strictly deterministic (i.e., all executors invalidate the same transactions), all executors produce the same output.

% We treat consensus as a Black-Box, where each node receives an identical chain of blocks $B_1, B_2, ..., B_n$ which is then handed to the execution engine in sequence. As long as the execution is strictly deterministic, this approach ensures that all server processes reach the same state. However, as execution does not occur on the critical path of consensus, client transactions cannot be fully validated before their inclusion in a block. Thus, we assume the output of consensus to be a \textit{dirty-ledger}~\cite{dirtyledger} $\mathcal{L}_i:= [tx_1,\dots,tx_i]$ where the resulting blockchain might contain invalid transactions (e.g., lacking funds or gas). Nonetheless, as long as the executors receive the same blocks in the same order (guaranteed by consensus) and the output of the execution of the chain of blocks is strictly deterministic (i.e., all executors invalidate the same transactions), all executors produce the same output. 

\begin{definition}[Valid Transition]
A correct node (i.e., executor or client) transitions from state \( S_i \) to \( S_{i+1} \) according to the transition function \( S_{i+1} = \delta(S_i, tx_i) \) if and only if:
\begin{enumerate}
    \item A quorum of at least \( 2f + 1 \) consensus nodes acknowledges transaction \(tx_i\).
    \item The transition \( \delta(S_i, tx_i) \) satisfies a validity predicate $\mathcal{P}$.

\end{enumerate}
Otherwise, if any of the above conditions is not met, the transaction is essentially ignored and \(S_{i+1} = S_i\).
\end{definition}
In the following, we assume that a state transition function \( \delta(S_i, tx_i) \) is deterministic i.e., there is a unique (if any) valid state $S_{i+1}$ given $S_i, tx_i$ and the validity predicate $\mathcal{P}$ is consistent among correct nodes, i.e., either a transition $\delta(S_i, tx_i)$ satisfies $\mathcal{P}$ in the view of every correct node, or it doesn't satisfy $\mathcal{P}$ in any correct node's view.
Consider an initial empty ledger \( \mathcal{L}_0 \) associated with the genesis state \( S_0 \). To determine the \( i \)-th state \( S_i \) of a ledger \( \mathcal{L}_i = [tx_1, tx_2, \dots, tx_i] \), where \( i > 0 \), transactions are applied iteratively as follows:
\(
S_i := \delta(\delta(\dots\delta(\delta(S_0, tx_1), tx_2) \dots), tx_i).
\)
For brevity, we denote:
\(
S_i := \delta(S_0, \mathcal{L}_i),
\)
as the successive application of all transactions \( tx \in \mathcal{L}_i \), starting from the initial state \( S_0 \).
\begin{definition}[Unconditional Validity]
A ledger \( \mathcal{L}_r \) at round \( r \) is said to be \textit{unconditionally valid} in the local view of a node iff given a predicate $\mathcal{P}$ the resulting state 
\(
S_r = \delta(S_0, \mathcal{L}_r)
\)
is computed through successive applications of \textit{valid} transitions.
\end{definition}

% Moving this over to overview
%To ensure determinism in the context of parallel execution, which is necessary to guarantee safety, we adopt a structure similar to Block-STM~\cite{blockstm}, where transactions go through two distinct phases. In the execution phase, transactions are executed optimistically, and in the validation phase, the execution read and write sets are cross-validated with the help of a multi-version data structure~\cite{mvdatastruct}. Validation and execution operate concurrently and if †inconsistencies are detected transactions are rolled back and rescheduled for execution. 

We adopt the standard Byzantine fault model required by the underlying partially synchronous BFT consensus algorithm. Concretely, among the \(N\) server processes, up to \(f < N/3\) may behave arbitrarily (Byzantine). For proof-of-stake blockchains, this translates to adversaries controlling less than $\frac{1}{3}$ of the total stake.
To maintain generality, references to a fraction of nodes in this paper can be interpreted as an equivalent fraction of stake in the context of hybrid proof-of-stake systems.

% Finally, we assume a Byzantine fault model consistent with the assumptions of the underlying consensus algorithm. Specifically, in most permissioned consensus algorithms operating in the partial synchrony model, in order to guarantee security properties the number of faulty nodes $f$ is bounded by $\frac{N-1}{3}$. For proof-of-stake blockchains, this translates to adversaries controlling less than $\frac{1}{3}$ of the total stake.
% To maintain generality, references to a fraction of nodes in this paper can be interpreted as an equivalent fraction of stake in the context of hybrid proof-of-stake systems.
% A given process is considered correct as long as it follows the protocol, otherwise, it is deemed faulty.
This upper bound on faulty processes ensures that consensus provides a ledger with the standard \emph{safety} and \emph{liveness} properties. If the bound is exceeded (e.g., a transient adversarial majority with \(f^\star \ge N/3\)), consensus safety or liveness may be violated. Nevertheless, \emph{unconditional ledger validity} still holds \emph{in the view of any correct executor}: because validators only order transactions and do not maintain application state, a correct executor advances its local state only via valid transitions (per \(\mathcal{P}\)) and therefore will not follow an invalid fork even when consensus finalizes an invalid block.
% This upper bound on the number of faulty nodes guarantees that the output of consensus is a ledger with the standard properties of \textit{safety} and \textit{liveness}. It is easy to see that even if these properties are violated when the number of faulty nodes exceeds the bound of $\frac{n}{3}$, \textit{unconditional ledger validity} still holds. Intuitively, this property protects correct nodes from an attacker who temporarily controls $f^*\geq N/3$ validators and attempts to fork the system into an arbitrary state. 
Note that such violations of the resilience threshold $f$ of the ledger are more likely to happen in systems that employ Delegated Proof of Stake (DPoS) such as Sui \cite{sui} and Cosmos, as the voting power is concentrated in a smaller number of nodes who do not risk their full collateral getting slashed in the case the system utilizes a fork detection or recovery algorithm.

\section{Design Overview}
\label{sec:overview}

The goal of \sys is to enable stragglers, i.e., lagging validators and clients, to catch up to the \emph{state} at the head of the chain by acting as executors, i.e., locally executing and validating blocks without depending on signed execution results, thereby upholding \emph{unconditional validity}. Drawing from our workload analysis and insights into the impact of block composition on performance, we establish specific design objectives for \sys. First, we want to speed up execution through careful transaction scheduling and avoid frequent re-executions without relaxing validity. Second, we want stragglers to be able to query and verify hints with minimal overhead.

Furthermore, as we will show in Section~\ref{sec:workloads}, the workloads we encounter in a blockchain setting are highly contended. Due to this, we anticipate high abort rates in optimistic execution engines, requiring frequent re-executions that slow down the execution. However, once some of the correct nodes finalized their execution, they can share execution insights with stragglers, such that stragglers can optimize scheduling and avoid re-executions to catch up to the head of the chain faster.

% To ensure determinism in the context of parallel execution, which is necessary to guarantee safety, we adopt a structure similar to Block-STM~\cite{blockstm}, where transactions go through two distinct phases. In the execution phase, transactions are executed optimistically, and in the validation phase, the execution read and write sets are cross-validated with the help of a multi-version data structure~\cite{mvdatastruct}. Validation and execution operate concurrently and if inconsistencies are detected transactions are rolled back and rescheduled for execution.
To ensure determinism under parallel execution we adopt a Block-STM–style design~\cite{blockstm} with two phases. In the execution phase, transactions run optimistically across cores, scheduled with the guidance of hints or without them. In the validation phase, transactions are checked in block order against a multi-version store~\cite{mvdatastruct}: each transaction’s reads must be consistent with the resulting state up to its position in that order. This check detects conflicts stemming from speculative execution. Conflicting transactions are rolled back and rescheduled; otherwise their writes commit. As a result, the committed outcome is equivalent to executing the transactions sequentially in block order.

\subsection{Hint Extraction and Storage}
In this section, we outline the process of extracting and propagating hints, and how nodes take advantage of them to catch up while still guaranteeing deterministic and correct execution.

\begin{figure}[t!]
\centering
\includegraphics[width= 0.2\linewidth]{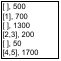}
\caption{Example Hint Structure for a Block with 6 Transactions}
\label{fig:hintstruct}
\end{figure}

\begin{algorithm}
\scriptsize
\centering
\caption{\textsl{Hint Extraction From Read/Write Sets}}
\label{algo:hintextract}
\begin{algorithmic}[1]
\State{$depmap \gets \emptyset$}
\State{$hintgraph \gets \emptyset$}
\Procedure{extract\_hints}{$readset, writeset, txid$}

\For{\textbf{each} $read \in readset$} \Comment{Iterate over reads}
    \If{$read \in depmap$}
        \State{$hintgraph_{txid} \gets hintgraph_{txid} \cup depmap_{read}$}
    \EndIf
\EndFor
\For{\textbf{each} $write \in writeset$} \Comment{Iterate over writeset}
    \If{$write \in depmap$}
        \State{$hintgraph_{txid} \gets hintgraph_{txid} \cup depmap_{write}$}
    \EndIf
    \State{$depmap_{write} \gets txid$}
\EndFor
\EndProcedure

\end{algorithmic}
\end{algorithm}

% \sys requires two types of hints for transaction scheduling. First, for each transaction $tx_i$, we need to know any preceding transaction $tx_j$ that modified any resource that $tx_i$ accesses. Furthermore, for each transaction $tx_i$, we also want to know its execution complexity.

\sys requires two kinds of hints for scheduling: (i) for each transaction \(tx_i\), the set of preceding transactions \(tx_j\) that modify resources accessed by \(tx_i\) (its dependencies); and (ii) an estimate of \(tx_i\)’s runtime (CPU time), i.e., how long it is expected to occupy a processing core.

This data can be obtained from the output of the optimistic execution on non-straggling nodes. For instance, in Block-STM, the multi-version data structure already tracks the read and write sets of all transactions, enabling data extraction outside the critical execution path. Similarly, the execution engine also already evaluates the execution complexity of each transaction during runtime, allowing this information to be retrieved without additional overhead.
As a result, \sys can be seamlessly integrated with any optimistic execution engine that allows extracting this data.

Fig.~\ref{fig:hintstruct} illustrates an example set of hints for a set of six transactions. Each row represents the hints for a specific transaction within a block. The first element in each row indicates the set of transactions the given transaction depends on (e.g., the second transaction depends on the first), while the second element represents the execution complexity.
This approach can be further optimized to minimize storage and bandwidth overhead. For instance, overlapping dependencies can be compressed. For three transactions $A,B,C$ If $B$ depends on $A$, and $C$ depends on both $A$ and $B$, it is sufficient to declare $B$ as a dependency for $C$.
\new{
This structure can be further compressed using efficient DAG compression algorithms~\cite{9006804}.
However, as shown in Section~\ref{sec:evaluation}, hint graphs constitute only a small fraction of the total block size, resulting in negligible storage and bandwidth overhead. In rare cases where a hint set is unusually large and would add non-trivial overhead, nodes may simply omit storing hints for that block. \sys is an optional acceleration layer: discarding hints does not affect consensus operation or the correctness of execution.

% In rare cases where a hint set exceeds the expected size and imposes non-trivial overhead, nodes may simply choose not to store hints for that block, as \sys is not critical to either safety or liveness.
}

Algorithm~\ref{algo:hintextract} illustrates this process, including hint compression. For each transaction, we iterate over the readset to record read dependencies in the $hintgraph$. Subsequently, we iterate over the writeset to record write dependencies and update the dependency map $depmap$ for the given resource with the last transaction that modified it.
This data can be stored alongside the block and disseminated to stragglers on request.

%The optimistic parallel execution engine Block-STM~\cite{blockstm} tracks all transactions with the help of a multi-version data structure that records the read- and write-sets of all transactions. This allows Block-STM to detect conflicts between concurrently executed transactions during run-time and enables it to initiate re-execution with fresher inputs when necessary.

%After deterministically finishing the execution of all transactions, the read and write-sets are already natively returned from the multi-version data structure as the resulting state from the parallel execution, alongside a footprint measured in gas that represents the complexity of the execution.

%This data can be stored alongside the block and disseminated to stragglers on request. To preserve bandwidth and storage, this data can be compressed by only tracking the closest dependency of a given transaction. For example, if transaction B depends on A and transaction C depends on both A and B, transaction C only needs to declare its dependency on B, as B already declared its dependency on A.

\subsection{Hint Propagation and Catching Up}

There are two ways how nodes may detect they are straggling. Either, by evaluating the progress of the other nodes in the network from the incoming execution state commit certificates as used for example in Aptos~\cite{aptos}, or by evaluating the number of blocks the node has queued up for execution. 

In either case, nodes can request execution hints to speed up their execution to catch up with the remaining nodes in the system.
In order to receive execution hints, stragglers can contact active nodes and request hints. Depending on how far behind a node is, it can obtain hints for several blocks or, if it is close to catching up, for a single block at a time.

Hints may be invalid in two ways: they may omit dependencies, resulting in rollback and transaction re-execution; or they may include spurious dependencies, which unnecessarily result in sequential execution and reduce performance. While \sys incorporates a validation mechanism similar to Block-STM~\cite{blockstm} to ensure correctness invalid hints can still degrade performance and cause stragglers to fall further behind, impacting overall system latency and throughput.

We now show that the worst-case overhead introduced by invalid hints is bounded. Due to the validation step, incorrect hints are detected immediately after executing the corresponding transaction. Thus, the first transaction with invalid hints triggers an immediate fallback to vanilla (optimistic) execution for the remaining transactions.

Let $\delta^v$ denote the time required to execute a block using vanilla execution, and $\delta^g$ the time under hint-guided execution. In the worst case, all but the hints for the last transaction are correct, and only the last transaction execution detects invalid hints and triggers a fallback. Then the total execution time is bounded by $\delta^e \leq \delta^v + \delta^g$. As we show in Section~\ref{sec:evaluation}, $\delta^g \leq \delta_v$, hence the execution overhead is tightly bounded by $2\delta^v$.

\ray{Furthermore, transactions that have been successfully executed and validated, provided that all preceding transactions in the block have also been executed and validated, do not have to be re-executed. This significantly reduces the re-execution overhead.}

Nonetheless, \ray{although the slowdown is bounded, a malicious validator can still obstruct a straggler from catching up}. As such, relying on a single validator to provide hints is undesirable. To address this, we introduce two operation modes: \textit{Optimistic Hints} and \textit{Deterministic Hints}.

\subsubsection{Optimistic Hints}

Since invalid hints cannot compromise safety in \sys, a straggler may request hints from a small, random, and geographically close subset of validators (e.g., $log(N)$) to reduce retrieval latency. If any of these hints prove invalid, the straggler broadcasts the signed message as proof of misbehavior and falls back to the deterministic mode. In the presence of a penalty or slashing mechanism, this evidence can be used to penalize the offending validators.

\subsubsection{Deterministic Hints}

Under normal operation, with at most $f$ faulty validators out of $N = 3f + 1$, a quorum of matching hints from $f + 1$ validators is guaranteed to be correct. This follows from the safety of the consensus protocol: the co-located \emph{executors} on all correct validators execute the same block deterministically and thus produce identical hints. Any matching set of $f + 1$ signed hints must include hints of at least one honest validator, who by definition only provides correct hints.

Conversely, if a straggler in deterministic mode receives invalid hints from a quorum of $f + 1$ validators, this indicates a violation of the adversarial threshold. In this scenario, where forks may occur, the straggler can broadcast the signed invalid hints as proof of misbehavior.
A potential consequence of this is the initiation of the recovery protocol~\cite{recoveryprotocols,roughgarden} as discussed in Section~\ref{sec:introduction}.

Finally, to preserve performance in low-contention scenarios, a correct node can signal that optimistic execution is preferable. This is achieved by replacing the usual hints with a single flag, thus bypassing the potential latency of guided execution when it offers no benefit.

In the decoupled design, validators that are behind on execution still participate in consensus but are more likely to propose stale transactions (e.g., insufficient gas). As such, narrowing this execution gap can improve the overall goodput of the system. Furthermore, when a fraction of transaction fees is distributed to all active validators, stale transactions may reduce the collective rewards. In this scenario, there is a clear incentive for validators to help their peers to catch up to the newest execution state, as it directly influences their rewards.

\section{Evaluation}
\label{sec:evaluation}

\subsection{Implementation}
\label{sec:implementation}

\begin{algorithm}[t!]
\scriptsize
\centering
\caption{\textsl{Transaction Scheduling}}
\label{algo:scheduletx}
\begin{algorithmic}[1]
\State{$queue \gets \emptyset$}
\State{$priorityqueue \gets \emptyset$}

\State{$depgraph \gets \emptyset$} \Comment{Parent/Child Relationship between transactions}
\State{$critpath \gets \emptyset$} \Comment{Graph describing the Critical Transaction path}

\Procedure{schedule}{$txn$, $hintgraph$}
\State{$depgraph \gets hintgraph$}
\For{\textbf{each} $tx \in txn$} \Comment{Iterate over Transactions}
    \If{$|depgraph_{tx}.parents| \leq 0$}
         \If{$|depgraph_{tx}.children| \leq 0$} \label{alg:line:queue1}
            \State{$queue \gets queue \cup tx$}
        \Else
            \State{$priorityqueue \gets priorityqueue \cup tx$}
        \EndIf
    \Else
        \State{$txcp \gets \bot$} \Comment{Track Critical Parent}
        \For{\textbf{each} $txp \in depgraph_{tx}.parents$} \Comment{Iterate Parent TX}
    \If{$txcp = \bot \lor txp.pathcost > txcp.pathcost$}
        \State{$txcp \gets txp$} \Comment{Add as Critical Parent}
    \EndIf
\EndFor
    \State{$critpath_{txcp} \gets critpath_{txcp} \cup tx$} \label{alg:line:prim}
    \EndIf
\EndFor
\EndProcedure

\Procedure{execute}{$tx$}
\For{\textbf{each} $txp \in depgraph_{tx}.parents$} \Comment{Iterate Parent TX}
    \If{$txp.status \neq Completed$}
        \State{$critpath_{txp} \gets critpath_{txp} \cup tx$} \Comment{Add as Critical Parent} \label{alg:line:crit}
        \State{\textsc{return}} \Comment{Don't execute}
    \EndIf
\EndFor
    \State{$tx.execute$} \Comment{Execute Transaction}
\For{\textbf{each} $txc \in critpath_{tx}$} \Comment{Iterate over critical children}
    \If{$|depgraph_{txc}.children| \leq 0$} \label{alg:line:queue2}
        \State{$queue \gets queue \cup tx$}
    \Else
        \State{$priorityqueue \gets priorityqueue \cup tx$}
    \EndIf
\EndFor
\EndProcedure

\end{algorithmic}
\end{algorithm}

To evaluate the feasibility of \sys, we implemented a guided execution engine on top of Block-STM~\cite{blockstm} in Aptos~\cite{aptos}. Furthermore, we implemented the hint extraction from the execution results\footnote{https://anonymous.4open.science/r/aptos-core-public-8F74}.

Our guided parallel execution utilizes the hints consisting of the dependency graph and the per-transaction execution time estimates to construct an efficient schedule that prioritizes computationally heavy dependency chains and minimizes costly re-executions. This scheduling logic is outlined in Algorithm~\ref{algo:scheduletx}.

\subsubsection*{Prioritizing critical parents}
The core of our strategy is to prioritize, for each dependent transaction, its \emph{critical parent}, i.e., the predecessor in the dependency graph with the highest cumulative execution cost. We implement this with a dual-queue system. Transactions without dependencies are scheduled immediately: those that are leaves in the dependency graph (i.e., have no children) are placed in a standard queue, while non-leaf transactions are placed in a priority queue (Line~\ref{alg:line:queue1}). For each non-source transaction, we select its critical parent and record that parent’s critical children (Line~\ref{alg:line:prim}). When a critical parent completes, its dependent children are \emph{optimistically} made eligible for execution (Line~\ref{alg:line:queue2}). This approach minimizes core idle time between dependent tasks, thereby enhancing system throughput.

\subsubsection*{Avoiding re-executions}
However, as this is done optimistically, before any transaction is executed it is necessary to verify that all parent transactions already finished executing before scheduling a transaction for execution; if any parent has not finished executing yet, the transaction is re-added as a child of this parent, and its execution is delayed (Line~\ref{alg:line:crit}). 

In parallel with this scheduling logic, we perform a validation step inspired by Block-STM~\cite{blockstm}. After a transaction executes, its read/write set is checked for conflicts against the write sets of preceding transactions using a multi-version data store~\cite{mvdatastruct}. A detected conflict, whether from speculative execution or an incorrect hint, triggers an immediate rollback and rescheduling of the offending transaction. For the remainder of the block, the system reverts to its standard optimistic execution model, and the peer that supplied the faulty hint is marked as untrustworthy. This validation process is highly efficient as it only queries metadata rather than spawning a virtual machine.

A key benefit of this guided approach is the reduction of speculative work. In highly contended workloads with long sequential paths, this can result in temporarily idle CPU cores. We leverage this opportunity for further optimization, such as using idle cores to verify transaction signatures, effectively moving this overhead off the critical path, as detailed in Section~\ref{sec:implementation}.

\subsubsection*{Optimizations}

Alongside this, we also added several optimizations that opened up due to the new scheduling approach. 
Traditionally, client transaction signatures are verified before handing the transaction to the execution engine.
In \sys, as the scheduler is aware of the dependency chains, we can leverage idle execution workers to verify client transaction signatures during execution, moving the signature verification away from the critical path. While Block-STM is unaware of the transaction dependency graph, idle workers could also be leveraged to verify client signatures. Therefore, to allow for a fair comparison we also implemented the signature verification optimization for Block-STM.

% Mention that this separation is a drawback for us, so, we gain back a bit of parallelism by doing this
% Mention, we only do this for the "next transaction" (only one user at a time).

Furthermore, in Aptos, transaction execution is split into three phases: i) prologue, ii) actual execution, and iii) epilogue. For each phase a virtual machine has to be instantiated, and, as such, for small transactions, each takes up roughly a third of the total execution time. As conflicts in the prologue and epilogue occur only when two concurrent transactions are from the same user, Block-STM almost fully optimistically parallelizes the prologue and epilogue, reducing the potential speedup \sys can achieve in practice. To compensate for this, we added additional functionality to \sys, to execute transaction prologues in parallel if they are the only or first transaction of a client in the block or the previous transaction of the client has been completed.

Next, to take advantage of cache locality, when a worker thread schedules the next set of child transactions, it selects one destined for the priority queue and places it at the top of its local queue.
Finally, we implemented the workloads from Section~\ref{sec:workloads} in Move Contracts and added code to benchmark the execution engine with and without hints.

\subsubsection*{Integrating \sys}

Integrating \sys into an existing blockchain is straightforward due to its modular design, requiring only minimal additions to the existing infrastructure. First, the extraction of hints, which is straightforward as the hints are a compressed version of the execution output. Second, hint storage, which can be short-term (e.g., limited to recent blocks for accelerating stragglers) or long-term (enabling new or recovering nodes to bootstrap by re-executing and re-validating historical blocks).
Next, nodes must detect when they fall behind, as described in Section~\ref{sec:overview}, and accordingly respond to hint requests.
Finally, \sys requires a guided execution engine that can accelerate execution using hints, while ensuring correctness even in the presence of incorrect hints.

\subsection{Benchmark}

We executed our experiments on a Debian GNU/Linux 12 server with a AMD EPYC 9654 96-Core Processor and 1024 GB of RAM. We created blocks of 10000 transactions with the help of our benchmarking suite and subsequently submitted them to Block-STM and \sys. Each benchmark consists of 9 configurations with between 2 and 32 worker threads respectively, where each worker thread maps to a dedicated core.
We executed each configuration a total of 10 times and then computed the average. 

We evaluate (i) the storage and bandwidth overhead of disseminating execution hints and (ii) the performance gains of \sys, indicating the speed-up stragglers can expect.

\paragraph*{\textbf{Execution Hints}}

The execution hints for each transaction contain a list of direct parent transactions and an execution complexity parameter in gas. We measured the overhead of the hints compared to the original transaction size and for all workloads, the resulting overhead is approximately 3\% of the raw transaction size.

\paragraph*{\textbf{Guided Parallel Execution}}

\begin{figure*}[ht]
\begin{center}
	\includegraphics[width=\linewidth]{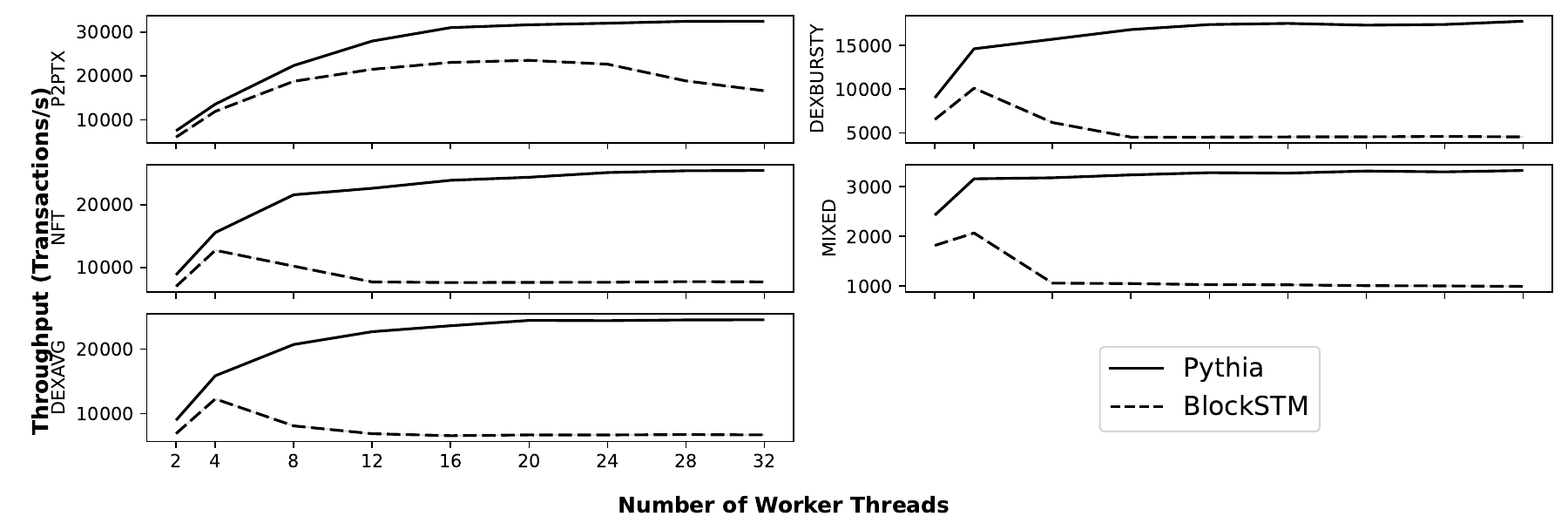}
\end{center}
\caption{Throughput per Second - Execution Engine}
\label{fig:evalallworkloads}
\end{figure*}

Figure ~\ref{fig:evalallworkloads} shows the per-second throughput for all workloads for the baseline Block-STM (dotted line) and the guided parallel execution \sys (solid line). The y-axis depicts the throughput in transactions per second and the x-axis the different configurations of worker threads from 2 to 32.

As a first observation, we note that \sys outperforms Block-STM across all configurations. It even guarantees a modest speedup for workloads with low levels of contention and achieves a more significant speedup as contention levels increase. The most consistent speedup is observed in the Mixed workload, where, in addition to high levels of contention, there is higher execution complexity. This makes re-execution more costly and shifts the balance of overhead from the prologue and epilogue to the actual execution.  

% Second, across all benchmarks, we observe that the performance benefit of increasing the number of worker threads eventually plateaus and, in the case of Block-STM, even begins to decline.
Second, across all benchmarks, we observe diminishing returns from adding additional worker threads. For \sys the performance plateaus, and for Block-STM it declines. This occurs because, as workload contention increases, the achievable parallelism reaches its limit. In contrast, \sys detects this limit and avoids further attempts to execute transactions in parallel. For native optimistic execution, however, adding more worker threads increases the abort rate, leading to more frequent re-executions.  

As a result, in any configuration, \sys allows stragglers to leverage hints to catch up with other nodes effectively. Moreover, the performance gains from \sys become more pronounced in computationally intensive and highly contended workloads. This is particularly critical, as nodes are more likely to struggle to keep up under these conditions.

% Should we add anything here?

\section{Related \& Future Work}
\label{sec:discussion}

% We divide related work into two categories. First,  benchmarks for blockchain execution engines, and second approaches to catch up in the context of parallel execution.

We group related work into three categories: (i) benchmarks for blockchain execution engines, (ii) approaches for catching up in the presence of parallel execution and (iii) guided parallel execution engines.

\paragraph*{\textbf{Execution Engine Benchmarks}}

Evaluating parallel smart contract execution engines requires benchmarks that accurately model real-world conditions, a need that existing tools do not fully meet. For example, while Diablo~\cite{diablo} is a comprehensive suite, its workloads (e.g., large data uploads or computationally intensive tasks) are not representative of typical blockchain transaction patterns and fail to capture realistic contention levels. Similarly, newer benchmarks like Blockbench~\cite{newbenchmark} draw from real-world applications but are tailored for sequential execution environments and lack the portability needed for cross-ecosystem comparisons. As a result, both are unsuitable for rigorously evaluating the effectiveness of parallel execution architectures.

% One of the most comprehensive benchmarks for blockchains is Diablo~\cite{diablo} which offers a full benchmark suite for blockchain performance evaluation.
% However, we identify several shortcomings. First, the workloads Diablo offers do not correspond to typical blockchain workloads (e.g., large data upload tasks or computationally intensive tasks). Furthermore, the chosen workloads neither correspond to typical usage patterns in terms of user distribution nor regarding contention levels. This makes these workloads unsuitable for evaluating the effectiveness of parallel smart contract execution engines.

% While newer approaches such as Blockbench~\cite{newbenchmark} offer benchmarks based on real-world blockchain workloads, they lack a focus on evaluating parallel execution engines with realistic levels of contention (i.e., they only evaluate traditional blockchains with sequential execution) and don't offer the same level of portability between eco-systems.

Our work addresses this gap by providing a microbenchmark specifically designed to reflect real-world data contention. We recognize that replaying historical transactions from sequential execution blockchains (e.g., Ethereum) can introduce artificial bottlenecks, as their smart contracts were not designed with concurrency in mind. To avoid this, we developed a set of simple smart contracts where conflicts are restricted to essential storage elements, thus capturing inherent rather than implementation-specific dependencies. Instead of a monolithic suite, our benchmark consists of a set of simple probability distributions paired with these simplified contracts. This approach ensures portability, allowing for easy integration into any testing framework and facilitating direct performance comparisons across different smart contract platforms (e.g., Ethereum and Solana Virtual Machines).

\paragraph*{\textbf{Catching Up Under Parallel Execution}}
Mechanisms for helping stragglers catch up in parallel execution blockchains often force a trade-off between security, usability, and performance. For instance, systems like Aptos~\cite{aptos} and Sui~\cite{sui} rely on signed checkpoints, forcing stragglers to trust execution results instead of independently verifying them. While this trust-based synchronization might be tolerable for bootstrapping a new node, \sys provides a superior alternative that allows even bootstrapping nodes to catch up quickly while fully validating the chain. However, for straggling validators the impact is significant: since they belong to the system’s failure model, skipping transaction execution and validity checks undermines overall security.

\paragraph*{\textbf{Guided Parallel Execution}}

Guided execution is a well-established approach in the parallel execution space. In systems like Solana~\cite{solana} and Sui~\cite{sui}, clients must provide strict access lists that enumerate every account or storage location a transaction is allowed to touch. These, in contrast to \sys, are not advisory hints but hard constraints; an overly conservative list serializes independent transactions, while an incomplete one causes the transaction to abort. This increases developer burden and can degrade parallelism: conservative access lists serialize otherwise independent work, whereas under-specification triggers failures or replays. Alternatives like Vegeta~\cite{vegeta} and Polygon’s design~\cite{polygonupdate} avoid client access lists by performing optimistic pre-execution on the consensus path. However, this introduces a new bottleneck, limiting overall throughput to the speed of the pre-execution phase~\cite{ethforward,polygonupdate}. Vegeta’s multi-leader design distributes this workload, but it keeps the proposer as the single point of failure for hint generation. Moreover, in a decoupled design it is difficult to distinguish between  an honest straggling proposer who may submit stale transactions due to lacking the correct state at the time of pre-execution from a malicious proposer attempting to degrade throughput. Furthermore, the resulting overhead from pre-execution is especially harmful to stragglers, who may be unable to propose blocks in time and fall further behind, degrading overall network performance. 

To the best of our knowledge, \sys is the first framework that enables stragglers to catch up in a modular blockchain environment without relaxing security guarantees. \sys bridges the optimistic and guided paradigms: the system executes optimistically by default, while stragglers can request hints to transition to a highly efficient guided-execution mode. It maximizes throughput by prioritizing long dependency chains and offloading tasks (e.g., signature verification) from the critical path to idle workers, all without imposing additional computational burden on clients or straggling validators.

Furthermore, \sys's modular architecture allows integration with guided engines that do not rely on hint accuracy for correctness. As future work, we envision applying this capability to platforms like Solana~\cite{solana} and Sui~\cite{sui}. Because client-provided access lists are often overly pessimistic, \sys could utilize an optimized dependency set obtained post-execution by executors to help stragglers synchronize faster. Realizing this would require substantial changes to those blockchains, which we leave for future work.

\section{Conclusion}
\label{sec:conclusion}

In this work, we presented \sys, a modular framework that leverages execution hints to allow straggling nodes and full nodes to catch up without relaxing the security model.
Depending on the workload \sys allows stragglers to execute blocks over 300\% faster.

We created a set of realistic workloads based on real-world data to help evaluate the effectiveness of parallel execution engines in practice. This allows comparing parallel execution engines of diverse ecosystems even if they use different virtual machines and smart contract languages.

As such, \sys opens the doors for a range of future work such as evaluating its impact in environments with guided parallel execution engines such as Solana~\cite{solana} or Sui~\cite{sui} where client hints might be overly pessimistic.

%Finally, our workload analysis has shown that the execution layer is still one of the major bottlenecks and hurdles to scaling blockchain systems, warranting further studies and novel approaches to alleviate this bottleneck.

\section{Acknowledgments}
This work was supported by the Austrian Science Fund (FWF) SFB project SpyCoDe F8502 and the Vienna Science and Technology Fund (WWTF) project SCALE2 CT22-045.

%-------------------------------------------------------------------------------
\section*{Availability}
%-------------------------------------------------------------------------------

The code of \sys as well as the real-world workloads and the data on which they are based will be made public upon acceptance of the paper.

%-------------------------------------------------------------------------------
\bibliographystyle{plain}
\bibliography{bib}

%%%%%%%%%%%%%%%%%%%%%%%%%%%%%%%%%%%%%%%%%%%%%%%%%%%%%%%%%%%%%%%%%%%%%%%%%%%%%%%%
\end{document}